\documentclass[11pt]{llncs}
\usepackage{graphicx}
\usepackage{amsmath}
\usepackage{amssymb}
\usepackage{hyperref}
\usepackage{url}
\usepackage[a4paper]{geometry}
\newtheorem{claim1}{Claim}

\renewcommand{\qed}{\hfill \ensuremath{\square}}
\newcommand{\qedTheorem}{\hfill \ensuremath{\blacksquare}}

\renewenvironment{proof}{
\vspace*{-\parskip}\noindent\textit{Proof.}}{$\qed$

\medskip
}

\newenvironment{proofTheorem}{
\vspace*{-\parskip}\noindent\textit{Proof.}}{$\qedTheorem$

\medskip
}

\makeatletter
\newsavebox\myboxA
\newsavebox\myboxB
\newlength\mylenA

\newcommand*\xoverline[2][0.75]{%
    \sbox{\myboxA}{$\m@th#2$}%
    \setbox\myboxB\null
    \ht\myboxB=\ht\myboxA%
    \dp\myboxB=\dp\myboxA%
    \wd\myboxB=#1\wd\myboxA
    \sbox\myboxB{$\m@th\overline{\copy\myboxB}$}
    \setlength\mylenA{\the\wd\myboxA}
    \addtolength\mylenA{-\the\wd\myboxB}%
    \ifdim\wd\myboxB<\wd\myboxA%
       \rlap{\hskip 0.5\mylenA\usebox\myboxB}{\usebox\myboxA}%
    \else
        \hskip -0.5\mylenA\rlap{\usebox\myboxA}{\hskip 0.5\mylenA\usebox\myboxB}%
    \fi}
\makeatother

\def\hexdigit#1{\ifnum#1<10 \number#1\else
\ifnum#1=10 A\else\ifnum#1=11 B\else\ifnum#1=12 C\else
\ifnum#1=13 D\else\ifnum#1=14 E\else\ifnum#1=15 F\fi
\fi\fi\fi\fi\fi\fi}
\font\tenmsam=msam10 at 11pt

\font\sevenmsam=msam8
\font\fivemsam=msam6
\newfam\msafam
\textfont\msafam=\tenmsam
\scriptfont\msafam=\sevenmsam
\scriptscriptfont\msafam=\fivemsam

\font\tenmsbm=msbm10 at 11pt

\font\sevenmsbm=msbm8
\font\fivemsbm=msbm6
\newfam\msbfam
\textfont\msbfam=\tenmsbm
\scriptfont\msbfam=\sevenmsbm
\scriptscriptfont\msbfam=\fivemsbm
\def\mathbb{\fam=\msbfam\tenmsbm}

\mathchardef\N="0\hexdigit\msbfam4E
\mathchardef\R="0\hexdigit\msbfam52
\mathchardef\Z="0\hexdigit\msbfam5A
\mathchardef\F="0\hexdigit\msbfam46
\mathchardef\le="3\hexdigit\msafam36
\mathchardef\ge="3\hexdigit\msafam3E
\def   \q     {\mbox{\quad}}

\def   \Mod   #1{\;{\rm mod}\;#1}
\def   \ceil  #1{\left\lceil#1\right\rceil}

\def  \beas        {\begin{eqnarray*}}
\def  \eeas        {\end{eqnarray*}}

\begin{document}

\title{Keyword-Based Delegable Proofs of Storage\thanks{A preliminary version of this work has been published in
International Conference on Information Security Practice and Experience (ISPEC 2018)
(available at \href{https://link.springer.com/chapter/10.1007/978-3-319-99807-7_17}{link.springer.com}).}}
\author{Binanda Sengupta \and Sushmita Ruj}
\index{Sengupta, Binanda}
\index{Ruj, Sushmita}
\institute{Indian Statistical Institute\\Kolkata, India\\
{\tt \{binanda\_r,sush\}@isical.ac.in}}

\maketitle

\pagestyle{plain}

\begin{abstract}
Cloud users (clients) with limited storage capacity at their end can outsource bulk data
to the cloud storage server. A client can later access her data by downloading the required data files.
However, a large fraction of the data files the client outsources to the server is often archival
in nature that the client uses for backup purposes and accesses less frequently.
An untrusted server can thus delete some of these archival data files in order to save some space
(and allocate the same to other clients) without being detected by the client (data owner).
\textit{Proofs of storage} enable the client to audit her data files uploaded to the server
in order to ensure the integrity of those files. In this work, we introduce a type of
(selective) proofs of storage that we call \textit{keyword-based delegable proofs of storage},
where the client wants to audit all her 
data files containing a specific keyword (e.g., ``important'').
Moreover, it satisfies the notion of \textit{public verifiability} where
the client can delegate the auditing task to a third-party auditor who audits the set of files
corresponding to the keyword on behalf of the client. We formally define the security of a
keyword-based delegable proof-of-storage protocol.
We construct such a protocol based on an existing
proof-of-storage scheme and analyze the security of our protocol. We argue that the techniques we use
can be applied atop \textit{any} existing publicly verifiable proof-of-storage scheme for static data.
Finally, we discuss the efficiency of our construction.

\end{abstract}

\section{Introduction}
Cloud computing platform provides a robust infrastructure to the cloud users (clients) in order to enable them storing
large amount of data on cloud servers. The clients can access their data as often as needed by downloading them from
the cloud servers.
Several storage service providers like Amazon Simple Storage Service (S3), Microsoft OneDrive, Dropbox and Google Drive
offer storage outsourcing facility to their clients (data owners). The clients pay these providers for the service
and expect that their (untampered) data can be retrieved at any point of time.
However, a client's data can be lost due to the failure of some of the storage nodes or due to a malicious activity
of the cloud server (an untrusted server can delete some part of the client's data in
order to save some space). Therefore, the client needs an assurance that her data files are stored by the server intact.
A possible cryptographic solution of the above problem is that the client computes an authenticator (tag) on a data file.
Then, she uploads the file and the tag to the server. To check the integrity of the data file, the client
downloads the file and the tag, and she checks if the file has been modified.
However, this solution is inefficient in practice
due to the large communication bandwidth required between the client and the cloud server.

\textit{Proofs of storage} provide an efficient mechanism to check the availability of the client's data
outsourced to a remote storage server.
In a proof-of-storage protocol, the client can \textit{audit} her data file stored on the server without accessing
the whole file, and still, be able to detect an unwanted modification of the file
done by the (possibly malicious) server.
Proof-of-storage protocols can be typically classified as:
\textit{provable data possession} (PDP) protocols~\cite{Ateniese_CCS,Erway_CCS,Wang_TPDS} and
\textit{proof-of-retrievability} (POR) protocols~\cite{JK_CCS,Wichs_ORAM,Stefanov_CCS}.
In these schemes, the client computes an authentication tag for each segment of her data file
and uploads the file along with these authentication tags.
Later, the client audits the data file via \textit{spot-checking},
where the client verifies the integrity of only a predefined number of randomly sampled segments of the file.
We note that PDP schemes provide the guarantee of retrievability
of \textit{almost all} segments of the data file.
On the other hand, \textit{all} segments of the file can be retrieved in POR schemes.
These schemes are designed for dynamic or static data
depending on whether the client can change the content of her data file after the initial outsourcing.
Some of these schemes are \textit{publicly verifiable} where anyone
with an access to the public parameters can perform audits.
In case of \textit{privately verifiable} schemes, only the client (data owner)
with some secret information can perform audits.
In a publicly verifiable proof-of-storage scheme, the client can \textit{delegate}
the auditing task to a third-party auditor (TPA) who performs audits on the client's data
and lets the client know if she finds any discrepancies.

The client often has a large repository of data files (documents), and she classifies these documents
for performing different types of analyses on them later.
Document-clustering is a popular technique where the data owner groups her data files depending on some
attributes of the files. Keyword-based document-clustering is one of the examples of document-clustering
where the clusters are formed
based on the distinct keywords present in the data files~\cite{Kang_2003,ChangH_IEICET05}.
It has various applications in data mining and information retrieval
such as designing an efficient scheme for searching  over these data files~\cite{GooglePatent_2014,GooglePatent_2017}.
Similarly, in a proof-of-storage protocol for such a clustered file repository, the client (data owner) might need
different degrees of availability-assurance for different outsourced files based on the keywords they contain.
Obviously, the client can check integrity of all the data files she has uploaded to the cloud server.
However, the cloud server can charge the client for the associated (computational and bandwidth) cost involved in an audit
(this cost is wasted in case the server is storing the client's data properly).
Typically, the more files are audited by the client, the higher is this associated cost.
Thus, the client might want to run audits only on important files having some specific keywords.
For example, the guarantee of availability for the files containing the keyword ``important''
might be of higher priority rather than that for the files containing the keyword ``movie''.
In this scenario, there must be some mechanism such that the client can efficiently check the integrity of all of her data files
(uploaded to the server) that contain a particular keyword.

We note that there exist many searchable encryption schemes~\cite{CurtmolaGKO06,CashJJKRS13,CashJJJKRS14} in the literature
that address efficient keyword-search over \textit{encrypted} data files stored on a remote server.
These schemes can be potentially integrated with existing proof-of-storage schemes
to audit the set of data files matching a particular keyword.
However, these searchable encryption schemes aim to minimize the information regarding the encrypted files that is
leaked to the (typically) semi-honest
remote server (that follows the protocol honestly but tries to learn some information regarding the content
of the files or the keywords being searched).
On the other hand, the untrusted server in proof-of-storage schemes is considered to be malicious
(i.e., it can corrupt the client's data in an arbitrary fashion).
Moreover, the definition of a proof-of-storage scheme does not demand encrypting the data files
or hiding the search (or access) patterns during a keyword-search (which involves storage and computational overhead).

\medskip
\noindent{\bf Our Contribution}\q
We summarize our contributions in this paper as follows.

\begin{itemize}
\item We introduce the notion of keyword-based delegable proofs of storage,
where the client can audit all her outsourced data files that contain a specific keyword (keyword-based audits).
Moreover, any third-party auditor with the knowledge of some public parameters can perform audits
on the set of files corresponding to the keyword on behalf of the client.\smallskip

\item We formalize the security model for a keyword-based delegable proof-of-storage protocol
and define the security for such a protocol.\smallskip

\item We construct a keyword-based delegable proof-of-storage protocol
based on an existing publicly verifiable proof-of-storage scheme and analyze the security of our protocol.
We describe a non-interactive challenge-generation method for keyword-based audits in our construction,
where the verifier does not know a priori the set of files matching a particular keyword.
%
Our techniques can be used with any existing publicly verifiable proof-of-storage scheme
for static data in order to enable keyword-based audits.\smallskip

\item We describe the efficiency of our keyword-based delegable proof-of-storage protocol.
\end{itemize}

The rest of the paper is organized as follows.
Section~\ref{prelim} describes some preliminaries and background related to our work.
In Section~\ref{overview:kdpos}, we introduce the notion of keyword-based delegable proofs of storage (KDPoS)
and discuss some possible constructions of a keyword-based delegable proof-of-storage protocol.
In Section~\ref{kdpos_scheme}, we propose a concrete KDPoS construction.
We define the security of a KDPoS scheme and analyze the security of our scheme in Section~\ref{sec:security}.
In Section~\ref{kdpos_efficiency}, we discuss the efficiency
of our KDPoS scheme.
In the concluding Section~\ref{conclusion}, we summarize the work done in this paper.

\section{Preliminaries and Background}
\label{prelim}

\subsection{Notation}
\label{notation}
Let $\lambda$ be the security parameter.
An algorithm $\mathcal{A}(1^\lambda)$ is a probabilistic polynomial-time
algorithm when its running time is polynomial in $\lambda$ and its output
is a random variable that depends on the internal coin tosses of $\mathcal{A}$.
An element $a$ chosen from a set $S$ uniformly at random
is denoted as $a\xleftarrow{R}S$.
A function $f:\N\rightarrow\R$ is called negligible in $\lambda$ if for
all positive integers $c$ and for all sufficiently large $\lambda$,
we have $f(\lambda)<\frac{1}{\lambda^c}$.
The notation `$\cdot||\cdots||\cdot$' denotes the concatenation of multiple strings.
For two integers $a$ and $b$ (where $a\leq b$), the set $\{a,a+1,\ldots,b\}$ is denoted by $[a,b]$ as well.

\subsection{Bilinear Maps}
\label{blmap}
Let $G_1,G_2$ and $G_T$ be multiplicative cyclic groups of prime order $p=\Theta(2^{2\lambda})$.
Let $g_1$ and $g_2$ be generators of the groups $G_1$ and $G_2$, respectively.
A bilinear map~\cite{KM_CC,Lynn_Thesis,Galbraith_DAM} is a function $e: G_1\times G_2\rightarrow G_T$ such that:

1. for all $u\in G_1, v\in G_2, a,b\in\Z_p$, we have $e(u^a,v^b)=e(u,v)^{ab}$
(bilinear property),

2. $e$ is non-degenerate, that is, $e(g_1,g_2)\not = 1$.

\noindent
Furthermore, properties 1 and 2 imply that

3. for all $u_1,u_2\in G_1, v\in G_2$, we have $e(u_1\cdot u_2,v)=e(u_1,v)\cdot e(u_2,v)$.

\noindent
If $G_1=G_2=G$, the bilinear map is symmetric; otherwise, it is asymmetric.
Unless otherwise mentioned, we consider bilinear maps which
are symmetric and efficiently computable. Let BLSetup$(1^\lambda)$ be
an algorithm which outputs $(p,g,G,G_T,e)$, the parameters of a
bilinear map, where $g$ is a generator of $G$ (i.e., $G=\langle g \rangle$).

\subsection{Discrete Logarithm Assumption}
\label{sec:DisLog}
The discrete logarithm problem over a multiplicative group $G=\langle g \rangle$ of prime order $p$ and generated by $g$
is defined as follows.
\begin{definition}[Discrete Logarithm Problem]
Given $g,h\in G$, the discrete logarithm problem over $G$ is to compute $a\in\Z_p$ such that $h=g^{a}$.
\end{definition}
We say that the discrete logarithm assumption holds in $G$ if, for any probabilistic polynomial-time
adversary $\mathcal{A}(1^\lambda)$, the probability
\beas
\Pr_{a\xleftarrow{R}\Z_p}[a\leftarrow\mathcal{A}(g,h):h=g^a]
\eeas
is negligible in $\lambda$, where the probability is taken over the internal
coin tosses of $\mathcal{A}$ and the random choices of $a$.

\subsection{Computational Diffie-Hellman Assumption}
\label{sec:CDH}
The computational Diffie-Hellman problem over a multiplicative group $G=\langle g \rangle$ of prime order $p$ and generated by $g$
is defined as follows.
\begin{definition}[Computational Diffie-Hellman Problem]
Given $g,g^a,h=g^b\in G$ for some $a,b\in\Z_p$, the computational Diffie-Hellman problem over $G$ is to compute $h^{a}\in G$.
\end{definition}
We say that the computational Diffie-Hellman assumption holds in the group $G$ if, for any probabilistic polynomial-time
adversary $\mathcal{A}(1^\lambda)$, the probability
\beas
\Pr_{a,b\xleftarrow{R}\Z_p}[h^{a}\leftarrow\mathcal{A}(g,g^a,h=g^b)]
\eeas
is negligible in $\lambda$, where the probability is taken over the internal
coin tosses of $\mathcal{A}$ and the random choices of $a$ and $b$.
We note that the hardness of the computational Diffie-Hellman problem in $G$ implies the hardness
of the discrete logarithm problem in $G$.

\subsection{Erasure Codes}
\label{sec:erasure_code}
A $(\tilde{m},\tilde{n},d)_\Sigma$-erasure code is an error-correcting code~\cite{MWSloane77}
that comprises an encoding algorithm Enc: $\Sigma^{\tilde{n}}\rightarrow\Sigma^{\tilde{m}}$
(encodes a message consisting of $\tilde{n}$ symbols into a longer codeword consisting of $\tilde{m}$ symbols) and
a decoding algorithm Dec: $\Sigma^{\tilde{m}}\rightarrow\Sigma^{\tilde{n}}$ (decodes a codeword to a message),
where $\Sigma$ is a finite alphabet and $d$ is the minimum distance
(Hamming distance between any two codewords is at least $d$) of the code.
The quantity $\frac{\tilde{n}}{\tilde{m}}$ is called the rate of the code.
A $(\tilde{m},\tilde{n},d)_\Sigma$-erasure code can tolerate up to $d-1$ erasures.
If $d=\tilde{m}-\tilde{n}+1$, we call the code a maximum distance separable (MDS) code.
For a $(\tilde{m},\tilde{n})$-MDS code, the original message can be reconstructed from any $\tilde{n}$
out of $\tilde{m}$ symbols of the codeword. Reed-Solomon codes~\cite{RSCode}
and their extensions are examples of non-trivial MDS codes.

\subsection{Bitcoin}
\label{sec:Bitcoin}
Nakamoto introduces a peer-to-peer cryptocurrency known as Bitcoin~\cite{Nakamoto08}
that does not rely on any trusted server.
The users in the Bitcoin network make payments by digitally signing transactions with their secret keys.
The network maintains a blockchain (a public ledger containing valid transactions) in a distributed fashion.
A new block is appended to the Bitcoin blockchain roughly in every 10 minutes (an epoch).
These blocks are generated by the miners (users trying to mine Bitcoins) in the network
who provide a cryptographic proof-of-work to show, in order to claim a mining reward, that
they have indeed expended a large amount of computational power.
Presently, Bitcoin uses Back's Hashcash~\cite{back02} as the proof-of-work.
The mining scheme in Bitcoin involves solving a cryptographic mining puzzle that is not precomputable.
Finding a solution of the puzzle works in the following way:
Let $T_1,T_2,\ldots,T_z$ be some of the valid transactions for a certain epoch
which are not included in any previous block. The miners try to find a nonce $\eta$ such that
SHA-256$(BH||root_{MHT}||\eta)\leq Z$,
where $Z$ is a predefined target value (the difficulty level),
$BH$ is the hash of the latest block appended to the Bitcoin blockchain
and $root_{MHT}$ is the root-digest of the Merkle hash tree built over $T_1,T_2,\ldots,T_z$.
Due to the preimage-resistance property of SHA-256, the only way to compute such a 
nonce $\eta$ is to search over all possible values of the nonce in a brute-force manner.

\subsection{Digital Signatures}
\label{dig_sig}
A digital signature scheme~\cite{GMR_ACM}
consists of the following polynomial-time algorithms:
a key generation algorithm KeyGen, a signing algorithm Sign
and a verification algorithm Verify. KeyGen takes as input the security parameter
$\lambda$ and outputs a pair of keys $(psk,ssk)$, where $ssk$ is the
secret key and $psk$ is the corresponding public verification key.
The algorithm Sign takes a message $m$ from the message space $\mathcal{M}$
and the secret key $ssk$ as input
and outputs a signature $\sigma$.
The algorithm Verify takes as input the public key $psk$, a message $m$
and a signature $\sigma$, and outputs \texttt{accept} or \texttt{reject}
depending upon whether the signature is valid or not.
Any of these algorithms can be probabilistic in nature.
The correctness and security (existential unforgeability under adaptive
chosen message attacks~\cite{GMR_ACM}) of a digital signature scheme are described as follows.
\begin{enumerate}
 \item \textit{Correctness}:\q Algorithm Verify always accepts a signature generated by an honest signer, that is,
	 \begin{align*}
	  \Pr[\text{Verify}(psk,m,\text{Sign}(ssk,m))=\texttt{accept}]=1.
	 \end{align*}
 \item \textit{Security}:\q Let Sign$_{ssk}(\cdot)$ be the signing oracle and $\mathcal{A}$
	 be any probabilistic polynomial-time adversary with an oracle access to Sign$_{ssk}(\cdot)$.
	 The adversary $\mathcal{A}$ adaptively makes polynomial (in $\lambda$)
	 number of sign queries to Sign$_{ssk}(\cdot)$ for different messages and gets back the signatures
	 on those messages. The signature scheme is secure if $\mathcal{A}$ cannot produce,
	 except with some probability negligible in $\lambda$,
	 a valid signature on a message not queried previously, that is, for
	 any probabilistic polynomial-time adversary $\mathcal{A}^{\text{Sign}_{ssk}(\cdot)}$,
	 the following probability
	 \beas
	  \Pr[(m,\sigma)\leftarrow\mathcal{A}^{\text{Sign}_{ssk}(\cdot)}(psk):
	  {m\not\in Q_{s}} \wedge \text{Verify}(psk,m,\sigma)=\texttt{accept}]
	 \eeas
	 is negligible in $\lambda$, where $Q_{s}$ is the set of queries made by
	 $\mathcal{A}$ to $\text{Sign}_{ssk}(\cdot)$.
\end{enumerate}

\subsection{Proofs of Storage}
Proofs of storage provide a client (data owner) with an efficient mechanism to verify
the integrity of her data outsourced to a remote cloud server.
Ateniese et al.~\cite{Ateniese_CCS} introduce the notion of \textit{provable data possession} (PDP)
where the client computes an authentication tag for each segment of her data file
and uploads the file along with these tags. During an audit, the client verifies the integrity of the
data file via \textit{spot-checking} where she samples
$l=O(\lambda)$ random segment-indices (challenge) and sends them to the server.
The server generates a proof based on the challenged segments (and their corresponding tags) and sends
the proof (response) to the client who verifies the proof.
This scheme also introduces the notion of ``public verifiability''
where the client can delegate the auditing task to a third-party auditor (TPA)
who performs audits on the client's behalf.
For a ``privately verifiable'' scheme, only the client with some secret information
can perform an audit.
Other schemes achieving PDP include~\cite{Erway_CCS,Wang_TPDS,Wang_TC,GrittiCSP17}.

Juels and Kaliski~\cite{JK_CCS} introduce the notion of \textit{proofs of retrievability} (POR) for static data.
The underlying idea of a proof-of-retrievability scheme is to encode the original file
with an erasure code (see Section~\ref{sec:erasure_code}),
authenticate the segments of the encoded file, and then upload
them on the cloud storage server~\cite{SW_JOC}. Due to the encoding, the server has to delete a large
number of segments to actually delete a file-segment, and this can be detected by the client with high probability.
This ensures that all segments of the file are retrievable from the responses
of the server which passes audits with some non-negligible probability.
Other POR schemes include~\cite{Bowers_CCSW,Wichs_HA,Xu_ASIACCS,Wichs_ORAM,Stefanov_CCS}.

\section{Keyword-Based Delegable Proofs of Storage}
\label{overview:kdpos}
In this section, we introduce the notion of keyword-based delegable proofs of storage
and describe some possible constructions.
We define a keyword-based delegable proof-of-storage protocol
as follows.
\begin{definition}\label{def:kdpos}
A keyword-based delegable proof-of-storage protocol (KDPoS)
consists of the following procedures.
\begin{itemize}
\item \emph{$\text{Setup}(1^\lambda)$:} The client runs this algorithm which sets the parameters of the protocol and
       generates a secret key-public key pair $K=(sk,pk)$ for the client.

\item \emph{$\text{Outsource}(\bar{F},sk,\xoverline{\texttt{fid}})$:} Given a set of data files $\bar{F}$ associated with
      a set of random file-identifiers \emph{\xoverline{\texttt{fid}}}, the client processes $\bar{F}$ to form
      another set of files $\bar{F}'$ (including respective authentication information computed using $sk$)
      and uploads $\bar{F}'$ to the server. The client stores some metadata $\bar{d}$ corresponding to $\bar{F}'$ at her end.
      
\item \emph{$\text{AuthRead}(j,\bar{F}',\bar{d},pk,\texttt{fid})$:} When the client wants to read the $j$-th block
      of a file $F'$ identified by \emph{\texttt{fid}}, the server sends to the client $F'[j]$, the $j$-th block of the file,
      along with the corresponding proof $\Pi(j)$.

\item \emph{$\text{VerifyRead}(j,\bar{d},pk,sk,F'[j],\Pi(j),\texttt{fid})$:} After receiving $(F'[j],\Pi(j))$
      from the server,
      the client checks the validity of $\Pi(j)$.
      The client outputs 1 if $\Pi(j)$ is a valid proof for $F'[j]$; she outputs 0, otherwise.
      
\item \emph{$\text{SChallenge}(pk,l,\bar{d},\widetilde{\texttt{fid}})$:} During an audit on a set of files identified by
      \emph{$\widetilde{\texttt{fid}}$} $\subseteq$ \emph{$\xoverline{\texttt{fid}}$}, the verifier\footnote{The verifier
      can be a third-party auditor (TPA) or the client (data owner) herself. In case the verifier is a TPA, the client shares
      the metadata $\bar{d}$ with the TPA.}
      sends the set \emph{$\widetilde{\texttt{fid}}$} and a random challenge set $Q$ of cardinality $l=O(\lambda)$ to the server.

\item \emph{$\text{SProve}(Q,pk,l,\bar{F}',\widetilde{\texttt{fid}})$:} Given the challenge set $Q$
      and a set of files identified by \emph{$\widetilde{\texttt{fid}}$}, the server computes a proof of storage $T$
      corresponding to the challenge set $Q$ and sends $T$ to the verifier.

\item \emph{$\text{SVerify}(Q,T,pk,l,\bar{d},\widetilde{\texttt{fid}})$:} The verifier checks
       whether $T$ is a valid proof of storage corresponding to the challenge set $Q$.
      The verifier outputs 1 if the proof passes the verification; she outputs 0, otherwise.

\item \emph{$\text{KChallenge}(pk,l,w,\bar{d})$:} During a keyword-based audit on the files containing a given keyword $w$, the verifier
      generates a token $t_w$ and a challenge set $Q$. The verifier sends $Q$ along with the token $t_w$ to the server.

\item \emph{$\text{KProve}(Q,pk,l,\bar{F}',t_w)$:} Upon receiving the challenge set $Q$, the server
      computes a proof of storage $T$ corresponding to all the data files containing the keyword $w$
      and sends $T$ to the verifier.

\item \emph{$\text{KVerify}(Q,T,pk,l,t_w,\bar{d})$:} The verifier checks
       whether $T$ is a valid proof of storage corresponding to the challenge set $Q$.
      The verifier outputs 1 if the proof passes the verification; she outputs 0, otherwise.

\end{itemize}
\end{definition}

We note that we have added two extra functionalities over a basic publicly verifiable proof-of-storage protocol
for static data: 
\textit{file-identifier-based audits} (audits on a set of files) and \textit{keyword-based audits}.
Definition~\ref{def:kdpos} implicitly includes a ``regular'' audit on a single data file associated with an identifier \texttt{fid},
where $\widetilde{\texttt{fid}}$ contains only the file-identifier \texttt{fid}.
A file-identifier-based audit consists of the procedures SChallenge, SProve and SVerify;
a keyword-based audit consists of the procedures KChallenge, KProve and KVerify.
An authenticated read comprises the procedures AuthRead and VerifyRead.

In order to perform a keyword-based audit for a particular keyword $w$, there are two options
that the client can adopt.
First, the client identifies the set (say, $F_w$) of files containing $w$ correctly
and runs a proof-of-storage protocol on these files. Second, the client sends to the server the keyword $w$
so that the server can identify the corresponding files itself and generate
responses during an audit properly.
From the security point of view, the client must be assured of the following guarantees.
\begin{enumerate}
 \item The integrity of the files outsourced to the (possibly malicious) server should be maintained, be it verified via file-identifier-based audits
 or via keyword-based audits. \smallskip
 
 \item For keyword-based audits, audits must be performed on \textit{all} the files containing the challenged keyword $w$
 (e.g., the server cannot cheat by sending proofs of storage for a set $F_w'\subsetneq F_w$
 or by sending proofs of storage for a set $F_w'\not\subset F_w$).
\end{enumerate}
Towards providing a framework of a keyword-based delegable proof-of-storage protocol, we describe briefly a series
of possible constructions and highlight some issues regarding each of these constructions.

\subsection{A Naive Approach}
The client builds an inverted index $\mathcal{I}$ on the set of files $\bar{F}$. A simple inverted index $\mathcal{I}$ stores,
for each keyword, a list of identifiers of all the files containing the keyword.
The client splits each (possibly encoded) file $F\in\bar{F}$ into segments
and generates
authentication tags on the segments using a secret key. She stores the inverted index
$\mathcal{I}$ at her end and uploads the files (and the tags) to the server. During an audit for a keyword $w$,
she consults $\mathcal{I}$ in order to get the identifiers of all files that contain $w$. With these file-identifiers known,
the client (or a TPA) can perform audits on the corresponding files and check their integrity.
However, this solution is not suitable for clients having low storage capacity (e.g., clients using low end mobile devices)
as the inverted index can be large (of the order of tens/hundreds of Gigabytes~\cite{CashJJJKRS14})
compared to the storage available at the client's side.

\subsection{Outsourcing Inverted Index to Storage Server}
To overcome the shortcomings of the naive solution, one possibility
is to outsource the inverted index $\mathcal{I}$ itself to the untrusted (malicious) storage server.
This enables the storage server to compute the responses given a keyword $w$.
Now, for a keyword-based audit, the verifier (the client herself or a TPA) sends the keyword $w$
along with a random challenge set $Q$ to the server.
The server searches in $\mathcal{I}$ for the identifiers of all files containing the specific keyword $w$,
computes a proof of storage for these files using $Q$ and sends the proof to the verifier (client or TPA).
However, the storage server in our security model (and in most of the existing proof-of-storage protocols)
is considered to be malicious, and it can delete a client's data in order to utilize the space thus gained
to store other clients' data (e.g., the server can actually store only 1,000 out of 10,000 files containing a particular keyword).
To be precise, there must be some mechanism for the verifier to check that:
1) the server is actually storing \textit{all} files containing the specific keyword and
2) the corresponding proofs of storage are computed \textit{only} on these files (during an audit).

\subsection{Our Approach}
\label{simpleApproach}

To achieve the guarantees mentioned above, we let the server store the inverted index (in the form of a lookup table)
in an authenticated fashion which makes the exact set of file-identifiers (matching a particular keyword)
returned by the server \textit{verifiable}.
Moreover, for each file outsourced to the server, the client embeds its file-identifier \texttt{fid}
in the authenticator tags computed on the segments of the file.
This ensures that the server responds with the proofs of storage computed on \textit{exactly} those files
that contain the particular keyword.

During an audit in a KDPoS scheme (see Definition~\ref{def:kdpos}), the verifier can perform either a file-identifier-based audit
(on a set of files) or a keyword-based audit (on the set of all files containing a particular keyword).
For a \textit{file-identifier-based audit}, the verifier selects a set of file-identifiers $\widetilde{\texttt{fid}}$ and generates a random challenge set $Q$
for $\widetilde{\texttt{fid}}$. Then, she sends $\widetilde{\texttt{fid}}$ and $Q$ to the server, and the server computes proofs of storage
based on these inputs. The verifier verifies the proofs with respect to $\widetilde{\texttt{fid}}$ and $Q$.

On the other hand, for a \textit{keyword-based audit}, the verifier cannot generate the random challenge set $Q$
for a keyword $w$ as she does not know a priori the set of file-identifiers $\widetilde{\texttt{fid}}$ matching $w$.
One trivial way to resolve this issue is the following.
The verifier sends the keyword $w$ to the server, and
the server sends the corresponding $\widetilde{\texttt{fid}}$ (with an authentication proof) to the verifier.
Then, the verifier generates $Q$ (using the procedure SChallenge) and sends it to the server.
However, this solution increases the number of communication rounds between the verifier and the server.
Another probable solution is to generate $Q$ in a non-interactive fashion such that
both the server and the verifier, given some randomness $\mathfrak{r}$, can produce the same challenge set $Q$.
However, this randomness $\mathfrak{r}$ used to generate $Q$ must be non-precomputable by the server
(and also verifiable by the verifier);
otherwise, a malicious server might manipulate $\mathfrak{r}$ to get $Q$ of its choice 
in order to pass a keyword-based audit.
We describe a non-interactive challenge-generation method used in our KDPoS construction
as follows.\bigskip

\noindent
\textbf{Non-interactive Challenge Generation for Keyword-Based Audits}\q
Armknecht et al.~\cite{Outpor_CCS} use a time-dependent pseudo-randomness
generator $\texttt{GetRandomness}: \mathcal{T}\rightarrow\{0,1\}^{l_{\texttt{seed}}}$
with an access to a secure time-dependent source, where $\mathcal{T}$
denotes a set of discrete timestamps and $l_{\texttt{seed}}=O(\lambda)$.
Let \texttt{cur} denote the current timestamp.
Given a timestamp $t\in\mathcal{T}$,
the generator \texttt{GetRandomness} outputs a uniform random string in $\{0,1\}^{l_{\texttt{seed}}}$
if $t\leq\texttt{cur}$; otherwise, it outputs $\bot$.
Armknecht et al.~instantiate \texttt{GetRandomness} by using Bitcoin (see Section~\ref{sec:Bitcoin}) as a secure time-dependent source
to achieve unpredictability of the output string.
For a timestamp $t\in\mathcal{T}$, \texttt{GetRandomness} outputs the
hash of the first block
appended to the Bitcoin blockchain after $t$. Given $t$, this pseudorandom output string can be generated (and verified) by anyone.
Although the original scheme~\cite{Outpor_CCS} uses this method in order to protect an honest party
(among the client, the cloud server and the third-party auditor) in case the other parties collude, this method
works well for non-interactive challenge generation  in our KDPoS scheme also (during keyword-based audits).

Given the pseudorandom string output by \texttt{GetRandomness},
the challenge set $Q$ can be generated in a similar way as described in~\cite{Miller14,Retricoin}.
The underlying idea is that the client sends to the server the current timestamp $t$ as a part of a challenge.
The server generates $Q$ based on $t$ and sends proofs of storage to the client. The client follows the same
procedure to generate $Q$ from $t$ and verifies the proofs sent by the server. We describe the method in details
for keyword-based audits (comprising the procedures KChallenge, KProve and KVerify) in our scheme.

\section{Our KDPoS Construction}
\label{kdpos_scheme}
We use the publicly verifiable POR scheme for static data proposed by Shacham and Waters~\cite{SW_JOC}
as the underlying proof-of-storage scheme, and modify the same in order to support keyword-based audits.
Our keyword-based delegable proof-of-storage protocol (KDPoS)
consists of the following procedures.

\begin{itemize}
\item $\textbf{Setup}(1^\lambda)$: Let the algorithm BLSetup$(1^\lambda)$ output $(p,g,G,G_T,e)$ as the parameters of a bilinear map,
      where $G$ and $G_T$ are multiplicative cyclic groups of prime order $p=\Theta(2^{2\lambda})$,
      $g$ is a generator of $G$ (i.e., $G=\langle g \rangle$) and
      $e:G\times G\rightarrow G_T$ (see Section~\ref{blmap}). The client chooses a random element $x\xleftarrow{R} \Z_p$ and
      sets $v=g^x$. Let $\alpha\xleftarrow{R} G$ be another generator of $G$ and
      $H:\{0,1\}^*\rightarrow G$ be the BLS hash~\cite{BLS_JOC} modeled as a random oracle~\cite{BR_RO}.
      Let $H_1:\{0,1\}^*\rightarrow \{0,1\}^{\ceil{\log_2 p}}$ be a (public) cryptographic hash function.
      Let $\mathcal{F}$ be the space of file-identifiers.
      Let $\texttt{GetRandomness}$ be a time-dependent pseudo-randomness generator as described in Section~\ref{simpleApproach}.
      For a given timestamp $t\in\mathcal{T}$, \texttt{GetRandomness} outputs the
      hash of the first block appended to the Bitcoin blockchain after $t$.
      Let $(ssk,psk)$ be the pair of signing and verification keys for
      a secure digital signature scheme $\mathcal{S}$ (see Section~\ref{dig_sig}).
      The secret key of the client is $sk=(x,ssk)$, the public key is $pk=(v,psk,\alpha)$.\medskip

\item $\textbf{Outsource}(\bar{F},sk,\xoverline{\texttt{fid}})$: Let the set of ${\bar{n}_f}$ data files
      the client wants to outsource to the server be $\bar{F}=\{F_1,F_2,\ldots,F_{\bar{n}_f}\}$.
      Let the file-identifiers corresponding to these files form the set
      $\xoverline{\texttt{fid}}=\{\texttt{fid}_1,\texttt{fid}_2,\ldots,\texttt{fid}_{\bar{n}_f}\}$,
      where each of these file-identifiers is drawn from the space $\mathcal{F}$ uniformly at random.
      The space $\mathcal{F}$ must be large enough (e.g., $\Z_p$) such that each file is associated with a distinct
      file-identifier except with a negligible probability.\medskip
      
      For each file $F\in\bar{F}$, the client extracts the keywords present in $F$. Let $W$ be the set of all \textit{distinct} keywords
      present in any of these files. The client builds a lookup table $T_{L}$ such that, for each keyword $w\in W$,
      the row indexed by the keyword $w$ contains an ordered list $L_w$ of file-identifiers matching $w$.
      For each row indexed by $w\in W$,
      the client computes a signature
      \begin{equation}\label{eqn:sigGen}
      \gamma_w=\mathcal{S}\text{.Sign}(ssk,w||L_w)
      \end{equation}
      and appends this signature $\gamma_w$ to $L_w$ present in that row.
      Let $n_w$ be the number of file-identifiers present in $L_w$. 
      Then, the row indexed by $w$ is of the form
      \[\begin{array}{|c|}
      \hline
      \texttt{fid}_{i_1}||\texttt{fid}_{i_2}||\cdots||\texttt{fid}_{i_{n_w}}||\gamma_w\\\hline
      \end{array}\]
      for some $i_1,i_2,\ldots,i_{n_w}\in[1,\bar{n}_f]$.
      We note that each signature $\gamma_w$ authenticates the binding between the \textit{exact} list ($L_w$) of file-identifiers matching $w$
      and the corresponding keyword $w$. \smallskip
      
      For each $i\in[1,\bar{n}_f]$, the client performs the following.
      \begin{itemize}
       \item The client encodes $F_i$ with an erasure code to form another file $F_i'$ with $n_i$ segments,
	     where $m_{ij}=F_i'[j]\in \Z_p$ for all $1\leq j\leq n_i$. \smallskip

       \item For all $1\leq j\leq n_i$, the client computes an authentication tag on the $j$-th segment as
	      \begin{equation}\label{eqn:tagGen}
		\sigma_{ij}=(H(\texttt{fid}_i||j)\cdot{\alpha}^{m_{ij}})^x\in G.
	      \end{equation}
	     Let $\Gamma_i=\{\sigma_{i1},\sigma_{i2},\ldots,\sigma_{in_i}\}$ be the ordered list of authentication tags for $F_i'$.
      \end{itemize}\smallskip
      
      Finally, the client uploads $\bar{F}'=(\{(F_i',\Gamma_i,\texttt{fid}_i,n_i)\}_{1\leq i\leq \bar{n}_f},T_L)$ to the cloud server.
      The client stores $\bar{d}=\{(\texttt{fid}_i,n_i)\}_{1\leq i\leq \bar{n}_f}$ at her end
      in order to check the integrity of some of these files later.\medskip
      
\item $\textbf{AuthRead}(j,\bar{F}',\bar{d},pk,\texttt{fid})$: When the client wants to read the $j$-th block
      of a file $F'$ identified by {\texttt{fid}}, the server sends to the client $F'[j]$, the $j$-th block of the file,
      along with its authentication tag $\sigma$.\medskip

\item $\textbf{VerifyRead}(j,\bar{d},pk,sk,F'[j],\sigma,\texttt{fid})$: After receiving the pair $(F'[j],\sigma)$
      from the server,
      the client checks whether
      \begin{equation}\label{eqn:verRead}
	\sigma\stackrel{?}=(H(\texttt{fid}||j)\cdot{\alpha}^{F'[j]})^x\in G.
      \end{equation}
      The client outputs 1 if  the equality holds; she outputs 0, otherwise.\medskip
      
\item $\textbf{SChallenge}(pk,l,\bar{d})$: During a file-identifier-based audit, the verifier selects an ordered list of file-identifiers $\widetilde{\texttt{fid}}$
      to be challenged.  For each $\texttt{fid}_i$ present in $\widetilde{\texttt{fid}}$, the verifier generates a random challenge set $Q_i=\{(r_j,\nu_j)\}_i$
      of cardinality $l=\lambda$,
      where each $r_j\xleftarrow{R}[1,n_i]$ and each $\nu_j\xleftarrow{R}\Z_p$.
      The verifier sends $\widetilde{\texttt{fid}}$ and $Q=\{\{(r_j,\nu_j)\}_i\}_{\texttt{fid}_i\in\widetilde{\texttt{fid}}}$ to the server.\medskip

\item $\textbf{SProve}(Q,pk,l,\bar{F}',\widetilde{\texttt{fid}})$: For each $\texttt{fid}_i\in\widetilde{\texttt{fid}}$,
      the server computes a pair $(\sigma_i,\mu_i)$,
      where
      \begin{equation}\label{eqn:proofGen1}
      \begin{aligned}
       \sigma_i & =\prod_{(r_j,\nu_j)\in Q_i}{\sigma_{ir_j}}^{\nu_j}\in G,\\
       \mu_i & =\sum_{(r_j,\nu_j)\in Q_i}\nu_jm_{ir_j}\Mod p\in\Z_p.
      \end{aligned}
      \end{equation}
      The server sends $T=\{(\sigma_i,\mu_i)\}_{\texttt{fid}_i\in\widetilde{\texttt{fid}}}$ to the verifier.\medskip

\item $\textbf{SVerify}(Q,T,pk,l,\bar{d})$: For each $\texttt{fid}_i\in\widetilde{\texttt{fid}}$,
      the verifier checks whether the equality
	\begin{equation}\label{eqn:proofVer1}
	  e(\sigma_i,g)\stackrel{?}=e\left(\prod_{(r_j,\nu_j)\in Q_i}H(\texttt{fid}_i||r_j)^{\nu_j}\cdot{\alpha}^{\mu_i},v\right)
	\end{equation}
      holds or not. The verifier outputs 1 if all the equalities hold; otherwise, she outputs 0.\medskip

\item $\textbf{KChallenge}(pk,l,w,\bar{d})$: During a keyword-based audit for a given keyword $w$, the verifier
      chooses two random strings $s_0$ and $s_1$ each of size $\lambda$ bits.
      She also chooses the current timestamp $t$.
      Finally, she constructs a token $t_w=w||s_0||s_1||t$. The challenge set $Q$ is \texttt{null}.
      The verifier sends the token $t_w$ to the server.\medskip

\item $\textbf{KProve}(Q,pk,l,\bar{F}',t_w)$: Initially, the challenge set $Q$ is \texttt{null}.
      The server parses the token $t_w$ as $w||s_0||s_1||t$.
      Given the keyword $w$, the server fetches $T_L[w]$
      containing the ordered list $L_w$ of matching file-identifiers and
      the corresponding signature $\gamma_w=\mathcal{S}\text{.Sign}(ssk,w||L_w)$.
      Given the timestamp $t$, the server computes the pseudorandom string $\texttt{str}_t=\texttt{GetRandomness}(t)$.\smallskip
      
      Let $\widetilde{\texttt{fid}}=L_w$.
      For each $\texttt{fid}_i\in\widetilde{\texttt{fid}}$, the challenge set $Q_i=\{(r_j,\nu_j)\}_i$ of cardinality $l=\lambda$ is generated as
	  \begin{equation*}
	    \begin{split}
	      \forall j\in \Z_l:\quad r_j &=H_1(\texttt{str}_t||\texttt{fid}_i||j||s_0)\Mod n_i + 1,\quad[\text{since }r_j\in\{1,2,\ldots,n_i\}]\\
				    \nu_j &=H_1(\texttt{str}_t||\texttt{fid}_i||j||s_1)\Mod p.
	    \end{split}
	\end{equation*}
      
      For each $\texttt{fid}_i\in\widetilde{\texttt{fid}}$,
      the server computes a pair $(\sigma_i,\mu_i)$,
      where
      \begin{equation}\label{eqn:proofGen2}
      \begin{aligned}
       \sigma_i & =\prod_{(r_j,\nu_j)\in Q_i}{\sigma_{ir_j}}^{\nu_j}\in G,\\
       \mu_i & =\sum_{(r_j,\nu_j)\in Q_i}\nu_jm_{ir_j}\Mod p\in\Z_p.
      \end{aligned}
      \end{equation}
      The server sends $T=(T_L[w],\{(\sigma_i,\mu_i)\}_{\texttt{fid}_i\in\widetilde{\texttt{fid}}})$ to the verifier.\medskip

\item $\textbf{KVerify}(Q,T,pk,l,t_w,\bar{d})$: Initially, the set $Q$ is \texttt{null}.
      The verifier parses $t_w$ as $w||s_0||s_1||t$ and $T$ as as $(T_L[w],\{(\sigma_i,\mu_i)\}_{\texttt{fid}_i\in L_w})$.
      She verifies the validity of the signature $\gamma_w$ by checking whether
      \begin{equation}\label{eqn:signVer}
	 \mathcal{S}\text{.Verify}(psk,w||L_w,\gamma_w)\stackrel{?}=\texttt{accept}.
      \end{equation}
      If the verification outputs \texttt{reject}, the verifier outputs 0.
      Otherwise, the verifier proceeds as follows.\smallskip
      
      Given the timestamp $t$, the verifier computes $\texttt{str}_t=\texttt{GetRandomness}(t)$.
      Let $\widetilde{\texttt{fid}}=L_w$.
      For each $\texttt{fid}_i\in\widetilde{\texttt{fid}}$, the challenge set $Q_i=\{(r_j,\nu_j)\}_i$ of cardinality $l=\lambda$ is generated as
	  \begin{equation*}
	    \begin{split}
	      \forall j\in \Z_l:\quad r_j  &=H_1(\texttt{str}_t||\texttt{fid}_i||j||s_0)\Mod n_i + 1,\\
				\nu_j &=H_1(\texttt{str}_t||\texttt{fid}_i||j||s_1)\Mod p.
	    \end{split}
	\end{equation*}
	
      For each $\texttt{fid}_i\in\widetilde{\texttt{fid}}$,
      the verifier checks whether the equality
	\begin{equation}\label{eqn:proofVer2}
	  e(\sigma_i,g)\stackrel{?}=e\left(\prod_{(r_j,\nu_j)\in Q_i}H(\texttt{fid}_i||r_j)^{\nu_j}\cdot{\alpha}^{\mu_i},v\right)
	\end{equation}
      holds or not. The verifier outputs 1 if all the equalities hold; otherwise, she outputs 0.\medskip

\end{itemize}

\noindent
\textbf{Observations}\q
We make the following observations regarding our KDPoS construction.
\begin{itemize}

\item We note that the verifier in our KDPoS construction can schedule an audit
in a future time when she is supposed to be offline. This can be done by setting the timestamp $t$
in future. 
We recall that, for a timestamp $t$, \texttt{GetRandomness} outputs the
hash of the first Bitcoin block
appended to the Bitcoin blockchain after $t$. 
Therefore, it is not possible for the server to compute the pseudorandom output of $\texttt{GetRandomness}(t)$
unless the current time $\texttt{cur}\ge t$. 
Thus, the server can generate
the challenge set $Q$ (which is derived from the output of $\texttt{GetRandomness}(t)$) only after time $t$.
On the other hand, if the verifier sends a random seed to the server for an audit scheduled in a future time $t$,
the (possibly malicious) server can compute the challenge set $Q$ (and also the proof of storage) immediately after getting the request, 
modify parts of the file and 
send the proof after the specified time $t$.

\item The random challenge set $Q$ used in file-identifier-based audits (involving the procedures SChallenge, SProve and SVerify)
can also be generated in a non-interactive fashion similar to that used in keyword-based audits.
This non-interactive (and verifiable) generation of random challenge set $Q$ reduces
the overall communication between the verifier and the server (as the verifier need not send $Q$ to the server).\smallskip

\item We have used techniques (such as building an authenticated lookup table $T_L$ over keywords present in data files, and 
generating the random challenge set $Q$ with the help of the Bitcoin blockchain) on top of a particular POR scheme~\cite{SW_JOC}
in order to construct a KDPoS scheme. 
We note that these techniques are independent of the underlying POR scheme and do not modify the same.
Thus, \textit{we can integrate our techniques with any existing publicly verifiable POR/PDP scheme for static data
(which is based on spot-checking random locations of a file) to obtain such a KDPoS construction}.
\end{itemize}

\noindent
\textbf{Correctness of Verification Eqn.~\ref{eqn:proofVer1} and Eqn.~\ref{eqn:proofVer2} (and Eqn.~\ref{eqn:signVer})}\q
For an honest server correctly storing all the challenged segments and their corresponding
authentication tags for a file identified by $\texttt{fid}_i$, we have
\begin{equation*} 
\begin{split}
\sigma_i 	& = \prod_{(r_j,\nu_j)\in Q_i}{\sigma_{ir_j}}^{\nu_j}\\
		& = \prod_{(r_j,\nu_j)\in Q_i}(H(\texttt{fid}_i||r_j)\cdot{\alpha}^{m_{ir_j}})^{\nu_jx}\\
		& = \left(\prod_{(r_j,\nu_j)\in Q_i}H(\texttt{fid}_i||r_j)^{\nu_j}\cdot\prod_{(r_j,\nu_j)\in Q_i}\alpha^{\nu_jm_{ir_j}}\right)^{x}\\
		& = \left(\prod_{(r_j,\nu_j)\in Q_i}H(\texttt{fid}_i||r_j)^{\nu_j}\cdot\alpha^{\sum_{(r_j,\nu_j)\in Q_i}\nu_jm_{ir_j}}\right)^{x}\\
		& = \left(\prod_{(r_j,\nu_j)\in Q_i}H(\texttt{fid}_i||r_j)^{\nu_j}\cdot\alpha^{\mu_i}\right)^{x}.
\end{split}
\end{equation*}

Substituting the value of $\sigma_i$ in $e(\sigma_i,g)$, we get
\begin{equation*} 
\begin{split}
e(\sigma_i,g) 	& = e\left(\left(\prod_{(r_j,\nu_j)\in Q_i}H(\texttt{fid}_i||r_j)^{\nu_j}\cdot\alpha^{\mu_i}\right)^{x},g\right)\\
		& = e\left(\prod_{(r_j,\nu_j)\in Q_i}H(\texttt{fid}_i||r_j)^{\nu_j}\cdot\alpha^{\mu_i},g^x\right)\\
		& = e\left(\prod_{(r_j,\nu_j)\in Q_i}H(\texttt{fid}_i||r_j)^{\nu_j}\cdot\alpha^{\mu_i},v\right).
\end{split}
\end{equation*}
On the other hand, correctness of Eqn.~\ref{eqn:signVer} directly follows from Eqn.~\ref{eqn:sigGen}
and the correctness property
of $\mathcal{S}$ (correctness of a digital signature scheme
is discussed in Section~\ref{dig_sig}).
Therefore, the proofs provided by an honest server always pass the verification Eqn.~\ref{eqn:proofVer1},
Eqn.~\ref{eqn:proofVer2} and Eqn.~\ref{eqn:signVer}.

\section{Security}
\label{sec:security}

A keyword-based delegable proof-of-storage protocol (KDPoS)
must satisfy the following properties.
The formal security definition of a KDPoS scheme is given later in this section.
\begin{enumerate}

\item \textbf{Authenticity}\q The authenticity requirements are twofold.
First, the cloud server must produce proofs of storage computed \textit{exactly} on the challenged files for file-identifier-based audits
(and proofs of storage computed \textit{exactly} on the files matching the challenged keyword in case of keyword-based audits).
Second, the cloud server cannot produce valid proofs during
audits without correctly storing the challenged segments of those files and their respective authentication information.\smallskip

\item \textbf{Retrievability}\q The retrievability property requires that, given a probabilistic
polynomial-time adversary $\mathcal{A}$ that can respond correctly to a challenge $Q$
with some non-negligible probability,
there exists a polynomial-time extractor algorithm $\mathcal{E}$ that can extract (at least)
the challenged segments of the challenged files (or the challenged segments of the files containing the challenged keyword)
by performing file-identifier-based audits (or keyword-based audits) with $\mathcal{A}$ for a polynomial (in $\lambda$) number of times.
The algorithm $\mathcal{E}$ has a rewinding access to $\mathcal{A}$.
The authenticity property restricts the adversary $\mathcal{A}$
to produce, during these interactions, valid responses (without storing these segments of the challenged files)
only with some probability negligible in $\lambda$.

\end{enumerate}

\subsection{Security Model}
\label{security_model}
We describe the following security game between the challenger (acting as the client)
and the adversary (acting as the cloud server).

\begin{itemize}

\item The challenger generates a secret key-public key pair and gives the public key to the adversary.
The adversary selects a set of files $\bar{F}$ associated with
a set of file-identifiers \xoverline{\texttt{fid}} to store.
The challenger processes $\bar{F}$ to form another set of files $\bar{F}'$ and returns $\bar{F}'$ to the adversary.
The adversary stores $\bar{F}'$ at its end.
The challenger stores only some metadata $\bar{d}$ for verification purpose.\smallskip

\item The adversary adaptively chooses and sends to the challenger a sequence of operations
defined by $\{\texttt{op}_i\}_{1\leq i\leq q}$ ($q$ is a polynomial in
the security parameter $\lambda$), where $\texttt{op}_i$ is an authenticated read or an audit.
The challenger executes these operations on the file stored by the adversary.
An audit request can be for either a file-identifier-based audit or a keyword-based audit,
where the set of files $\widetilde{\texttt{fid}}\subseteq\xoverline{\texttt{fid}}$
or the keyword is chosen by the adversary.
For each audit, the challenger executes an audit on the designated files stored by the adversary
(using SChallenge, SProve and SVerify for a file-identifier-based audit, or
using KChallenge, KProve and KVerify for a keyword-based audit).
The challenger lets the adversary know the result of each verification
(i.e., the output of VerifyRead or the output of SVerify or the output of KVerify).\smallskip

\item Let $\bar{F}^*$ be the final state of the set of files initially outsourced to the adversary after $q$ operations.
Finally, the challenger executes an audit protocol (file-identifier-based or keyword-based) with the adversary as follows.
The challenger chooses a set of file-identifiers (or a token for a keyword chosen by the challenger) and
sends them along with a random challenge set $Q$ to the adversary,
and the adversary returns a cryptographic proof to the challenger.
The adversary wins the game if it passes the verification.

\end{itemize}

\begin{definition}[Security of a KDPoS Scheme]\label{def:sec_kdpos}
A keyword-based delegable proof-of-storage protocol
is secure if, given any probabilistic polynomial-time
adversary $\mathcal{A}$ who can win the security game mentioned above with some non-negligible
probability, there exists a polynomial-time extractor algorithm $\mathcal{E}$ that can extract,
except with some probability negligible in $\lambda$,
(at least) the challenged segments of the files (that are challenged via file-identifier-based/keyword-based audits)
by interacting with $\mathcal{A}$ polynomially many times.
\end{definition}

\subsection{Security Analysis of Our KDPoS Scheme}
\label{sec_analysis}

We have described the security model of a keyword-based delegable proof-of-storage protocol (KDPoS)
in Section~\ref{security_model}. In this section, we state and prove the following theorem in order to prove that our
KDPoS construction described in Section~\ref{kdpos_scheme} is secure in this security model.
Our scheme is secure in the random oracle model~\cite{BR_RO} under the computational Diffie-Hellman assumption over $G=\langle g \rangle$
(see Section~\ref{sec:CDH}).
We note that the fraction of the file-segments retrievable by the extractor algorithm $\mathcal{E}$
interacting with the adversary $\mathcal{A}$ (in Definition~\ref{def:sec_kdpos}) depends on the underlying
proof-of-storage scheme used. If the underlying scheme is a PDP scheme, $\mathcal{E}$ can retrieve
\textit{at least} the challenged file-segments. On the other hand, if the underlying scheme is a POR scheme
(as in our KDPoS construction), $\mathcal{E}$ can retrieve \textit{all} the segments of the files.

\begin{theorem}\label{theorem_kdpos}
Given that the computational Diffie-Hellman assumption holds in $G$ and
the underlying digital signature scheme is secure, the KDPoS scheme described in Section~\ref{kdpos_scheme}
is secure in the random oracle model according to Definition~\ref{def:sec_kdpos}.
\end{theorem}

\begin{proofTheorem}
We use the following claim in order to prove Theorem~\ref{theorem_kdpos}.

\begin{claim1}\label{claim_authenticity_kdpos}
Given that the computational Diffie-Hellman assumption holds in $G$ and the underlying
digital signature scheme is secure, the authenticity of the challenged file-segments is guaranteed in the random oracle model.
\end{claim1}

\begin{proof}
We prove that the following properties required for authenticity hold in our KDPoS scheme.
First, the adversary $\mathcal{A}$ must produce proofs of storage computed \textit{exactly}
on the challenged files for file-identifier-based audits (and proofs of storage computed \textit{exactly}
on the files matching the challenged keyword in case of keyword-based audits).
Second, $\mathcal{A}$ cannot produce valid proofs during
audits without correctly storing the challenged segments of those files and their respective authentication information.
We prove these two properties in Part I and Part II, respectively.\medskip

\noindent
\textbf{Part I}\q
During a keyword-based audit for a given keyword $w$, let the correct content of the row $T_L[w]$
be the ordered list $L_w$ of file-identifiers matching $w$ and
the corresponding signature $\gamma_w=\mathcal{S}\text{.Sign}(ssk,w||L_w)$.
Now, if the adversary $\mathcal{A}$ can produce a valid signature $\gamma_w'=\mathcal{S}\text{.Sign}(ssk,w||L_w')$
for a different list $L_w'$ ($\not =L_w$) of file-identifiers for the same keyword $w$,
it breaks the security of the signature scheme
(see Section~\ref{dig_sig} for the security definition of a signature scheme).
However, as the underlying signature scheme $\mathcal{S}$ used in our KDPoS scheme is assumed to be secure,
$\mathcal{A}$ can forge such a signature only with a negligible probability.
Thus, similar to a file-identifier-based audit, the challenger knows the \textit{exact} file-identifiers
on which the keyword-based audit is being performed. We note that, for each file stored by the adversary,
the unique file-identifier is embedded in the authentication tags corresponding to all segments
of that file (see Eqn.~\ref{eqn:tagGen} in Section~\ref{kdpos_scheme}). Therefore,
once the challenger knows the exact file-identifiers on which an audit (keyword-based/file-identifier-based)
is being performed, it is not possible for the adversary to generate proofs of storage computed
on other files without being detected, except with a negligible probability
(this follows from the proof described in Part II).\medskip

\noindent
\textbf{Part II}\q
We refer~\cite{SW_JOC} for the detailed proof. Here, we provide a brief sketch of the same.
The basic idea of the proof is as follows.
We assume that, if possible, the adversary $\mathcal{A}$ produces valid proofs (i.e., they pass the verification) during
an audit without correctly storing the challenged segments and their respective authentication information.
Then, we can construct another probabilistic polynomial-time algorithm $\mathcal{B}$ which can
solve problems that are assumed to be hard in $G=\langle g \rangle$. Without loss of generality, we assume that $\mathcal{A}$
produces a valid (but incorrect) proof $(\sigma'_i,\mu'_i)$ for a particular file $\texttt{fid}_i\in\widetilde{\texttt{fid}}$
challenged with $Q_i=\{(r_j,\nu_j)\}_i$. Let $(\sigma_i,\mu_i)$ be the correct proof (i.e., computed honestly) corresponding to
the same challenge set $Q_i=\{(r_j,\nu_j)\}_i$ for $\texttt{fid}_i$. Now, we consider the following two cases
based on whether $\sigma_i$ is equal to $\sigma'_i$.\medskip

\begin{itemize}
\item[] \textbf{Case I:}\q
Let $\sigma_i\not=\sigma'_i$.
As both of the proofs $(\sigma_i,\mu_i)$ and $(\sigma'_i,\mu'_i)$ pass the verification, we have
\begin{equation}\label{eqn:security1}
 e(\sigma_i,g)=e\left(\prod_{(r_j,\nu_j)\in Q_i}H(\texttt{fid}_i||r_j)^{\nu_j}\cdot{\alpha}^{\mu_i},v\right)
\end{equation}
and 
\begin{equation}\label{eqn:security2}
 e(\sigma'_i,g)=e\left(\prod_{(r_j,\nu_j)\in Q_i}H(\texttt{fid}_i||r_j)^{\nu_j}\cdot{\alpha}^{\mu'_i},v\right).
\end{equation}
We observe that if $\mu_i=\mu'_i$, then $\sigma_i=\sigma'_i$. So, $\mu_i\not=\mu'_i$. We define $\Delta_{\mu}=\mu'_i-\mu_i\not=0$.

Let the algorithm $\mathcal{B}$ be given with $g,g^a,h=g^b\in G$ for some $a,b\xleftarrow{R}\Z_p$. The goal of $\mathcal{B}$ is to compute $h^{a}\in G$
(i.e., to break the computational Diffie-Hellman assumption over $G$ described in Section~\ref{sec:CDH}).
$\mathcal{B}$ sets the public key $v=g^a$ and simulates the random oracle $H$ as follows.
For any random oracle query made by the adversary $\mathcal{A}$, the algorithm $\mathcal{B}$ chooses a random $\tilde{r}\xleftarrow{R}\Z_p$
and responds with $g^{\tilde{r}}\in G$. For queries of the form $H(\texttt{fid}_i||u)$,
the algorithm $\mathcal{B}$ chooses random values $\beta,\phi\xleftarrow{R}\Z_p$ and sets $\alpha=g^\beta\cdot h^\phi$
(during preprocessing the file identified by $\texttt{fid}_i$ as per $\mathcal{A}$'s request).
For each $u\in[1,n_i]$, $\mathcal{B}$ chooses a random $\tilde{r}_u\xleftarrow{R}\Z_p$,
sets $H(\texttt{fid}_i||u)=g^{\tilde{r}_u}/(g^{\beta m_{iu}}\cdot h^{\phi m_{iu}})$ and 
computes $\sigma_{iu}=(H(\texttt{fid}_i||u)\cdot\alpha^{m_{iu}})^a=(g^a)^{\tilde{r}_u}$.

Dividing Eqn.~\ref{eqn:security2} by Eqn.~\ref{eqn:security1}, we get
\begin{align*}
	     & e(\sigma'_i/\sigma_i,g) = e\left({\alpha}^{\Delta_{\mu}},v\right)\\
    \implies & e(\sigma'_i\cdot\sigma_i^{-1},g) = e\left({g}^{\beta\Delta_{\mu}}\cdot {h}^{\phi\Delta_{\mu}},v\right)\\
    \implies & e(\sigma'_i\cdot\sigma_i^{-1},g) = e\left({g},v^{\beta\Delta_{\mu}}\right)\cdot e\left({h}^{\phi\Delta_{\mu}},v\right)\\
    \implies & e(\sigma'_i\cdot\sigma_i^{-1}\cdot v^{-\beta\Delta_{\mu}},g) = e\left({h},v\right)^{\phi\Delta_{\mu}}\\
    \implies & e(\sigma'_i\cdot\sigma_i^{-1}\cdot v^{-\beta\Delta_{\mu}},g) = e\left({h}^a,g\right)^{\phi\Delta_{\mu}}\q[\text{since }v=g^a]\\
    \implies & h^a = (\sigma'_i\cdot\sigma_i^{-1}\cdot v^{-\beta\Delta_{\mu}})^{\frac{1}{\phi\Delta_{\mu}}}.
   \end{align*}
Now, the value of $\phi\Delta_{\mu}$ is zero with probability $1/p$ that is negligible in the security parameter $\lambda$
(as $\Delta_{\mu}\not=0$ and $\phi\xleftarrow{R}\Z_p$).
Thus, $\mathcal{B}$ breaks the computational Diffie-Hellman assumption over $G$.\medskip

\item[] \textbf{Case II:}\q
From Case I, we know that $\sigma_i$ must be equal to $\sigma'_i$.
We define $\Delta_{\mu}=\mu'_i-\mu_i\not=0$ (otherwise, if $\mu_i=\mu'_i$, then two proofs are equal).
Let the algorithm $\mathcal{B}$ be given with $g,h=g^{a}\in G$, where $a\xleftarrow{R}\Z_p$.
The goal of $\mathcal{B}$ is to compute $a$
(i.e., to break the discrete logarithm assumption over $G$ described in Section~\ref{sec:DisLog}).
When $\mathcal{A}$ requests for storing the file identified by $\texttt{fid}_i$,
the algorithm $\mathcal{B}$ chooses random values $\beta,\phi\xleftarrow{R}\Z_p$
and computes the tags taking $\alpha$ as $g^\beta\cdot h^\phi$.
Since $\sigma_i=\sigma'_i$, we get
$$e\left(\prod_{(r_j,\nu_j)\in Q_i}H(\texttt{fid}_i||r_j)^{\nu_j}\cdot{\alpha}^{\mu_i},v\right) =
e\left(\prod_{(r_j,\nu_j)\in Q_i}H(\texttt{fid}_i||r_j)^{\nu_j}\cdot{\alpha}^{\mu'_i},v\right)$$
from Eqn.~\ref{eqn:security1} and Eqn.~\ref{eqn:security2}.
This implies
	     ${\alpha}^{\mu_i} = {\alpha}^{\mu'_i}
    \implies 1 = {\alpha}^{\Delta_{\mu}}
    \implies 1 = (g^\beta\cdot h^\phi)^{\Delta_{\mu}}
    \implies h = g^{-\frac{\beta\Delta_{\mu}}{\phi\Delta_{\mu}}}$.
Now, the value of $\phi\Delta_{\mu}$ is zero with probability $1/p$ that is negligible in the security parameter $\lambda$
(as $\Delta_{\mu}\not=0$ and $\phi\xleftarrow{R}\Z_p$).
Finally, $\mathcal{B}$ outputs $a={-\frac{\beta\Delta_{\mu}}{\phi\Delta_{\mu}}}$ as the discrete logarithm of $h$ in $G=\langle g \rangle$
(thus, $\mathcal{B}$ breaks the discrete logarithm assumption over $G$).
\end{itemize}
We note that the hardness of the computational Diffie-Hellman problem in $G$ implies the hardness of the discrete logarithm problem in $G$.
This completes the proof of Claim~\ref{claim_authenticity_kdpos}.
\end{proof}

We describe the extraction procedure for a single encoded file $F'_i$ identified by $\texttt{fid}_i$.
The procedure works for each challenged file identified by
$\texttt{fid}\in\widetilde{\texttt{fid}}$.\footnote{For a file-identifier-based challenge, the list $\widetilde{\texttt{fid}}$ of file-identifiers
is already known. On the other hand, for a keyword-based challenge, the correct list $\widetilde{\texttt{fid}}$ of file-identifiers
matching a particular keyword is obtained initially (see Part I of Claim~\ref{claim_authenticity_kdpos}).
Then, file-identifier-based challenges are used to extract all the segments of each file identified by $\texttt{fid}\in\widetilde{\texttt{fid}}$.
So, the extraction procedures for these two types of challenges are almost same.}
We define a polynomial-time extractor algorithm $\mathcal{E}$ that can extract
all segments of $F'_i$ (except with negligible probability) by interacting with an adversary $\mathcal{A}$
that wins the security game described in Section~\ref{security_model} with some non-negligible probability.
As our KDPoS scheme satisfies the \textit{authenticity} property,
the adversary $\mathcal{A}$ cannot produce a valid proof $(\sigma_i,\mu_i)$
for a given challenge set $Q_i$ without storing
the challenged segments of $F'_i$ and their corresponding tags properly,
except with some negligible probability (see Claim~\ref{claim_authenticity_kdpos}).
This means that if the verification procedure outputs 1 during the extraction phase,
$\mu_i$ is indeed the correct linear combination of the untampered segments
(that is, $\mu_i=\sum_{(r_j,\nu_j)\in Q_i}\nu_jm_{ir_j}\Mod p$).

Suppose that the extractor algorithm $\mathcal{E}$ wants to extract
$k$ segments (indexed by $J$) of the file $F'_i$. It challenges $\mathcal{A}$ with $Q_i=\{(r_j,\nu_j)\}_{r_j\in J}$.
If the proof is valid (that is, if the verification procedure outputs 1), $\mathcal{E}$ initializes a matrix
$M_\mathcal{E}$ with $[\nu_{1r_j}]_{r_j\in J}$ as its first row, where $\nu_{1r_j}=\nu_{j}$ for each $r_j\in J$.
The extractor challenges $\mathcal{A}$ for the same $J$ but with different random coefficients.
If the verification procedure outputs 1 and the vector of coefficients is linearly independent to
the existing rows of $M_\mathcal{E}$, then $\mathcal{E}$ appends this vector to $M_\mathcal{E}$ as a row.
The extractor algorithm $\mathcal{E}$ runs this procedure until the matrix $M_\mathcal{E}$ has $k$ number of
\textit{linearly independent rows}.
So, the final form of the \textit{full-rank} matrix $M_\mathcal{E}$ is $[\nu_{ur_j}]_{u\in[1,k], r_j\in J}$.
Consequently, the challenged segments can be extracted using Gaussian elimination.

Following the way mentioned above, the extractor algorithm $\mathcal{E}$ can interact with $\mathcal{A}$
(polynomially many times) in order to extract $\rho$-fraction of segments
(for some $\rho$) present in the file $F'_i$ by setting the index set $J$ appropriately.
Use of a $\rho$-rate erasure code ensures retrievability of all segments of $F'_i$.
This completes the proof of Theorem~\ref{theorem_kdpos}.
\end{proofTheorem}

\section{Efficiency of Our KDPoS Scheme}
\label{kdpos_efficiency}
The efficiency of our KDPoS scheme depends on the underlying POR scheme~\cite{SW_JOC}.
For each file challenged by the verifier, the proof consists of
a pair of the form $(\sigma,\mu)$ that is of size $2\cdot\log_2 p$ bits, 
where $\sigma\in G$, $\mu\in\Z_p$ and $p=\Theta(2^{2\lambda})$.
For example, such a pair is 64 bytes long for 128-bit security (i.e., $\lambda=128$).
On the other hand, for each challenged file, the verifier needs to compute 2 pairings
along with other operations ($l+1$ exponentiations and one multiplication in $G$).
However, we later describe a method in order to make both of these parameters
independent of the number of files being audited (see Section~\ref{BatchPairing}).

\subsection{Storage Overhead}
We have described our KDPoS scheme assuming that an authentication tag (an element of $G$) is generated for each segment (an element of $\Z_p$) of a file.
Therefore, the storage overhead (for the tags) is same as the storage itself. This can be mitigated by grouping $s$ segments as a single chunk and
computing an authentication tag for each of these chunks~\cite{SW_JOC}. Thus, the storage overhead is ${1}/{s}$-fraction of the storage.
However, during an audit, the size of the aggregated segment ($\mu$) sent by the server as a proof is now $s\cdot\log_2 p$ bits.
In addition, we have introduced an authenticated lookup table $T_L$ in order to enable keyword-based audits.
Let $W$ be the set of all distinct keywords present in any file $F\in\bar{F}$
and $n_w$ be the number of files containing a keyword $w\in W$.
Let $b_{\mathcal{S}}$ be the size (in bits) of a signature in $\mathcal{S}$
and $b_{\mathcal{F}}$ be the number of bits required to represent the space of file-identifiers $\mathcal{F}$.
Then, the storage overhead incurred for storing $T_L$ is
$\sum_{w\in W}(n_w\cdot b_{\mathcal{F}}+b_{\mathcal{S}})$ bits.

\subsection{Efficient Search over Lookup Table} 
In order to enable the server to search over the lookup table $T_L$ efficiently,
the client builds a dictionary data structure (e.g., trie, hash table, self-balancing binary search tree)
over all distinct keywords in $W$ (during the procedure Outsource).
The node in this data structure corresponding to a keyword $w$ contains $L_w||\gamma_w$,
where $L_w$ is the ordered list of file-identifiers matching $w$ and $\gamma_w=\mathcal{S}\text{.Sign}(ssk,w||L_w)$.
Then, the client uploads this data structure
along with the processed files to the server.
Now, the server makes an efficient search to find the exact node corresponding to the keyword
challenged during a keyword-based audit.

\subsection{Communication Complexity}
In case of a keyword-based audit, the server sends to the verifier the row of $T_L$ indexed by $w$ (containing the list of file-identifiers matching $w$)
along with the corresponding proofs of storage (during the procedure KProve).
We note that for a file-identifier-based audit as well, the verifier sends a list of file-identifiers
along with the challenge set (during the procedure SChallenge).
Therefore, the overall communication complexity is of the same order for both types of audits.
Moreover, the challenge set $Q$ can be generated in a non-interactive way (for both types of audits) eliminating
the need for communicating the same.

\subsection{Proof Generation and Verification in a Batch}
\label{BatchPairing}
We observe that given a list of file-identifiers $\widetilde{\texttt{fid}}$ and
a challenge set $Q=\{\{(r_j,\nu_j)\}_i\}_{\texttt{fid}_i\in\widetilde{\texttt{fid}}}$,
the server computes a pair $(\sigma_i,\mu_i)$ for each $\texttt{fid}_i\in\widetilde{\texttt{fid}}$
(see Eqn.~\ref{eqn:proofGen1} and Eqn.~\ref{eqn:proofGen2}). Therefore, the corresponding proof $T$ consists of
$|\widetilde{\texttt{fid}}|$ pairs of the form $(\sigma,\mu)$, where $\sigma\in G$ and $\mu\in\Z_p$.
Hence, the proof size is $|T|=2\cdot|\widetilde{\texttt{fid}}|\cdot\log_2 p$ bits.
On the other hand, for each $\texttt{fid}_i\in\widetilde{\texttt{fid}}$, the verifier checks whether the following equality
 \begin{equation}\label{eqn:proofVer3}
  e(\sigma_i,g)\stackrel{?}=e\left(\prod_{(r_j,\nu_j)\in Q_i}H(\texttt{fid}_i||r_j)^{\nu_j}\cdot{\alpha}^{\mu_i},v\right)
 \end{equation}
holds or not (see Eqn.~\ref{eqn:proofVer1} and Eqn.~\ref{eqn:proofVer2}). Therefore, the verifier needs to perform
expensive pairing operations for $2\cdot|\widetilde{\texttt{fid}}|$ times. \medskip

To reduce both the size of the proof and the number of pairing operations required to verify a proof,
we adopt an idea similar to that of aggregating the challenged segments and their corresponding tags for a single file.
We observe that, for each challenged file identified by $\texttt{fid}_i$ in our KDPoS scheme,
the server aggregates all the challenged segments into a single segment $\mu_i$ and
all the corresponding tags into a single tag $\sigma_i$ (see Eqn.~\ref{eqn:proofGen1} and Eqn.~\ref{eqn:proofGen2}),
and the verifier runs the verification procedure
on the aggregated segment and the aggregated tag (see Eqn.~\ref{eqn:proofVer3}).
We extend this simple idea for multiple files (that are present in $\widetilde{\texttt{fid}}$) as follows.
During the proof generation procedure, the server computes a pair $(\sigma,\mu)$, where
$$\sigma=\prod_{\texttt{fid}_i\in\widetilde{\texttt{fid}}}\prod_{(r_j,\nu_j)\in Q_i}{\sigma_{ir_j}}^{\nu_j}=\prod_{\texttt{fid}_i\in\widetilde{\texttt{fid}}}\sigma_i\in G$$
and
\begin{align*}
\mu  =\sum_{\texttt{fid}_i\in\widetilde{\texttt{fid}}}\sum_{(r_j,\nu_j)\in Q_i}\nu_jm_{ir_j}\Mod p
     =\sum_{\texttt{fid}_i\in\widetilde{\texttt{fid}}}\mu_i\Mod p\in\Z_p.
\end{align*}
Given the aggregated segment $\mu$ and the aggregated tag $\sigma$, the verifier checks if
 \begin{equation}\label{eqn:proofVerBatch}
  e(\sigma,g)\stackrel{?}=e\left(\prod_{\texttt{fid}_i\in\widetilde{\texttt{fid}}}\prod_{(r_j,\nu_j)\in Q_i}H(\texttt{fid}_i||r_j)^{\nu_j}\cdot{\alpha}^{\mu},v\right).
 \end{equation}

In this case, the reduced proof size is $|T|=2\cdot\log_2 p$ bits,
and the verifier needs to perform only 2 pairing operations.
It is important to note that both of these parameters are now constant,
irrespective of the number of files involved in either a file-identifier-based audit
or a keyword-based audit.

\noindent
\textbf{Correctness of Verification Eqn.~\ref{eqn:proofVerBatch}}\q
For an honest server storing all the challenged segments and their corresponding
authentication tags (for each file identified by $\texttt{fid}_i\in\widetilde{\texttt{fid}}$) correctly, we have
\begin{equation*} 
\begin{split}
\sigma 	& = \prod_{\texttt{fid}_i\in\widetilde{\texttt{fid}}}\prod_{(r_j,\nu_j)\in Q_i}{\sigma_{ir_j}}^{\nu_j}\\
		& = \prod_{\texttt{fid}_i\in\widetilde{\texttt{fid}}}\prod_{(r_j,\nu_j)\in Q_i}(H(\texttt{fid}_i||r_j)\cdot{\alpha}^{m_{ir_j}})^{\nu_jx}\\
		& = \left(\prod_{\texttt{fid}_i\in\widetilde{\texttt{fid}}}\prod_{(r_j,\nu_j)\in Q_i}H(\texttt{fid}_i||r_j)^{\nu_j}\cdot\prod_{\texttt{fid}_i\in\widetilde{\texttt{fid}}}\prod_{(r_j,\nu_j)\in Q_i}\alpha^{\nu_jm_{ir_j}}\right)^{x}\\
		& = \left(\prod_{\texttt{fid}_i\in\widetilde{\texttt{fid}}}\prod_{(r_j,\nu_j)\in Q_i}H(\texttt{fid}_i||r_j)^{\nu_j}\cdot\prod_{\texttt{fid}_i\in\widetilde{\texttt{fid}}}\alpha^{\sum_{(r_j,\nu_j)\in Q_i}\nu_jm_{ir_j}}\right)^{x}\\
		& = \left(\prod_{\texttt{fid}_i\in\widetilde{\texttt{fid}}}\prod_{(r_j,\nu_j)\in Q_i}H(\texttt{fid}_i||r_j)^{\nu_j}\cdot\prod_{\texttt{fid}_i\in\widetilde{\texttt{fid}}}\alpha^{\mu_i}\right)^{x}\\
		& = \left(\prod_{\texttt{fid}_i\in\widetilde{\texttt{fid}}}\prod_{(r_j,\nu_j)\in Q_i}H(\texttt{fid}_i||r_j)^{\nu_j}\cdot\alpha^{\sum_{\texttt{fid}_i\in\widetilde{\texttt{fid}}}{\mu_i}}\right)^{x}\\
		& = \left(\prod_{\texttt{fid}_i\in\widetilde{\texttt{fid}}}\prod_{(r_j,\nu_j)\in Q_i}H(\texttt{fid}_i||r_j)^{\nu_j}\cdot\alpha^{\mu}\right)^{x}.
\end{split}
\end{equation*}

Substituting the value of $\sigma$ in $e(\sigma,g)$, we get
\begin{equation*} 
\begin{split}
e(\sigma,g) 	& = e\left(\left(\prod_{\texttt{fid}_i\in\widetilde{\texttt{fid}}}\prod_{(r_j,\nu_j)\in Q_i}H(\texttt{fid}_i||r_j)^{\nu_j}\cdot\alpha^{\mu}\right)^{x},g\right)\\
		& = e\left(\prod_{\texttt{fid}_i\in\widetilde{\texttt{fid}}}\prod_{(r_j,\nu_j)\in Q_i}H(\texttt{fid}_i||r_j)^{\nu_j}\cdot\alpha^{\mu},g^x\right)\\
		& = e\left(\prod_{\texttt{fid}_i\in\widetilde{\texttt{fid}}}\prod_{(r_j,\nu_j)\in Q_i}H(\texttt{fid}_i||r_j)^{\nu_j}\cdot\alpha^{\mu},v\right).
\end{split}
\end{equation*}

\section{Conclusion}
\label{conclusion}
In this work, we have introduced keyword-based delegable proofs of storage,
where the data owner (or a third-party auditor) can selectively check the integrity of all her data files containing a particular keyword.
We have formally defined the security of a keyword-based delegable proof-of-storage protocol.
We have provided a construction of an efficient keyword-based delegable proof-of-storage protocol
and analyzed the security of our construction. 
Any existing publicly verifiable proof-of-storage scheme (based on spot-checking techniques) can be extended in a similar fashion
as described in this work.
We have also discussed the efficiency of our construction and some possible ways to enhance this efficiency.

\section*{Acknowledgments}
This work is partially supported by Cisco University Research Program Fund, 
CyberGrants ID: \#698039 and Silicon Valley Community Foundation. 
The authors would like to thank Chris Shenefiel and Samir Saklikar for their comments and suggestions.


\end{document}